\documentstyle[prb,aps,epsf,eqsecnum]{revtex}
\begin{document}
\draft
\twocolumn[\hsize\textwidth\columnwidth\hsize\csname  
@twocolumnfalse\endcsname
\author{P. Pieri and G.C. Strinati}
\address{Dipartimento di Fisica, UdR INFM, 
Universit\`{a} di Camerino, I-62032 Camerino, Italy}
\title{Scattering length for composite bosons in the BCS-BEC crossover}

\maketitle

\begin{abstract}
The present manuscript concerns the calculation of the boson-boson scattering 
length for the composite bosons that form as bound-fermion pairs in the 
strong-coupling limit
of the BCS-BEC crossover. The material presented in this manuscript was
already published as a part of a longer paper on the BCS-BEC crossover problem 
(P. Pieri and G.C. Strinati, Phys. Rev. B {\bf 61}, 15370 (2000)).
Given the recent experimental advances on the formation of ultracold
bosonic molecules from a Fermi gas of atoms by a Feshbach resonance,
the calculation of the boson-boson scattering length has now become 
of particular interest.  
The present short version of the above paper could thus be helpful to 
the scientific community working on ultracold atomic gases.
Accordingly, the present manuscript is intended for circulation as a preprint 
only. 
\end{abstract}
\hspace*{-0.25ex}
]
\vspace{0.8cm}
\narrowtext

The recent formation of ultracold bosonic molecules from a Fermi
gas of atoms\cite{regal} by a Feshbach resonance allows for an experimental 
check of theoretical calculations for physical quantities within the 
BCS-BEC crossover. 
In particular, in Ref.~\onlinecite{PS00} the relation between the 
composite-boson scattering length $a_B$ and the fermionic scattering length 
$a_F$ was calculated in the strong-coupling limit of the BCS-BEC crossover.   
The summation therein of all bosonic T-matrix diagrams has led to the result
$a_B=0.75 a_F$. This result corrects the value $a_B= 2 a_F$  
obtained within the Born approximation for the effective residual bosonic 
interaction\cite{Haussmann,PS-96,epjb}. 
The result $a_B=0.75 a_F$ could be tested experimentally in the near future, 
by measuring at the same time the molecule-molecule scattering length and
the fermion-fermion scattering length while scanning the magnetic field through
the Feshbach resonance.

In this manuscript, we provide a condensed version of the material 
published in Ref.~\onlinecite{PS00}, 
focusing specifically on the calculation of the 
bosonic scattering length.
We hope that this short summary of our previous work could be 
useful to the scientific community at the present time.

\section{Building blocks of the diagrammatic structure for composite bosons}

In this section, we discuss the diagrammatic structure that describes 
generically the composite bosons in terms of the constituent fermions.
Our theory rests on a judicious choice of the fermionic interaction, 
which (without loss of generality) greatly reduces the number and 
considerably simplifies the expressions of the Feynman diagrams to be taken 
into account. 

\subsection{Regularization of the fermionic interaction}

We begin by considering the following Hamiltonian for interacting
fermions (we set Planck $\hbar$ and Boltzmann $k_{B}$ constants equal to unity
throughout):
\begin{eqnarray}
& & H  =  \sum_{\sigma} \int d{\bf r} \, \psi_{\sigma}^{\dagger}({\bf r}) \left(
- \frac{\nabla^2}{2m} - \mu \right) \psi_{\sigma}({\bf r})  \nonumber  \\
& & +  \frac{1}{2} \sum_{\sigma, \sigma'} \int d{\bf r} \, d{\bf r'}
  \psi_{\sigma}^{\dagger}({\bf r}) \psi_{\sigma'}^{\dagger}({\bf r'})
  V_{{\mathrm eff}}({\bf r}-{\bf r'}) 
  \psi_{\sigma'}({\bf r'}) \psi_{\sigma}({\bf r}) . 
                                                         \label{Eq:Hamiltonian}
\end{eqnarray}
Here, $\psi_{\sigma}({\bf r}$) is the fermionic field operator with spin 
projection 
$\sigma = (\uparrow, \downarrow$), $m$ the fermionic mass, $\mu$ the 
fermionic chemical potential, and $V_{{\mathrm eff}}({\bf r}-{\bf r'})$ the 
{\it effective potential\/} that provides the \emph{attraction\/} between 
fermions.
For the application to atomic gases, the two spin states correspond to two
different hyperfine states of the fermionic atoms.

To simplify the ensuing many-body diagrammatic structure 
(yet preserving the physical effects we are after), we adopt for 
$V_{{\mathrm eff}}$ the form of a ``contact'' potential \cite{footnote-contact}
\begin{equation}
V_{{\mathrm eff}}({\bf r}-{\bf r'}) = v_{0} \,\, \delta ({\bf r}-{\bf r'})
                                         \label{Eq:deltafunc}
\end{equation}
where $v_{0}$ is a negative constant. 
With this choice, the interaction affects only fermions with opposite spins in
the Hamiltonian (\ref{Eq:Hamiltonian}) owing to Pauli principle.
A suitable \emph{regularization\/} of the potential (\ref{Eq:deltafunc}) is, however, 
required to get accurate control of the many-body diagrammatic structure. 
In particular, the equation (in the center-of-mass frame)
\begin{equation}
\frac{m}{4 \pi a_{F}} \, = \, \frac{1}{v_{0}} \, + \, \int \! \frac{d{\bf k}}
{(2 \pi)^{3}} \frac{m}{{\bf k}^{2}}                    \label{ferm-scatt-ampl}
\end{equation}
for the \emph{fermionic scattering length\/} $a_{F}$ associated with the 
potential (\ref{Eq:deltafunc}) is ill-defined, since the integral over the 
three-dimensional wave vector ${\bf k}$ is ultraviolet divergent.
The delta-function potential (\ref{Eq:deltafunc}) is then regularized, by 
introducing an ultraviolet cutoff $k_{0}$ in the integral of Eq.~(\ref{ferm-scatt-ampl}) and letting
$v_{0} \, \rightarrow \, 0$ as $k_{0} \, \rightarrow \, \infty$, in order to
keep $a_{F}$ fixed at a \emph{finite\/} value.
The required relation between $v_{0}$ and $k_{0}$ is obtained directly from 
Eq.~(\ref{ferm-scatt-ampl}). One finds:
\begin{equation}
v_{0} \, = \, - \, \frac{2 \pi^{2}}{m k_{0}} \, - \, \frac{\pi^{3}}{m a_{F} k_{0}^{2}}
                                                               \label{vo}
\end{equation}
when $k_{0} |a_{F}| \gg 1$.  

\begin{figure}
\narrowtext
\epsfxsize=3.3in 
\epsfbox{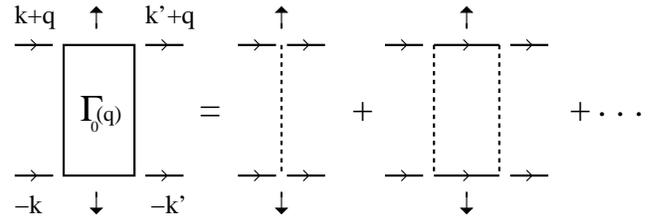}
\vspace{.2truecm}
\caption{ Particle-particle ladder. 
Four-momenta are indicated and spin labels are represented by
up and down arrows. Dotted lines represent the interaction potential and full 
lines the fermionic bare single-particle Green's functions.}
\label{fig:pplad}
\end{figure}

With the regularization (\ref{vo}) for the potential, the classification of the
many-body diagrams gets considerably simplified, since only specific 
sub-structures of these diagrams survive when the limit 
$k_{0} \, \rightarrow \, \infty$ is eventually taken. In particular, 
in order to obtain a finite result for a given Feynman diagram,
the vanishing strength $v_0$ of the potential should be compensated by an 
ultraviolet divergence in some internal wave-vector integration.   
For the particle-particle ladder of 
Fig.~\ref{fig:pplad}, the internal wave-vector integration associated with 
each 
rung diverges in the limit $k_0\to\infty$ and compensates the vanishing of
$v_0$, yielding the finite result:
\begin{eqnarray}
& &\Gamma_{0}(q) = - \left\{ \frac{m}{4 \pi a_{F}} + \right.
\int \! \frac{d{\bf k}}{(2 \pi)^{3}}\nonumber\\ 
& &\times\left. \left[\frac{\tanh(\beta \xi({\bf k})/2) 
+\tanh(\beta \xi({\bf k-q})/2)}{2(\xi({\bf k})+\xi({\bf k-q})-i
\Omega_{\nu})} 
- \frac{m}{{\bf k}^{2}} \right] \right\}^{-1}\; .
\label{p-p ladder}
\end{eqnarray}
Here, $\xi({\bf k}) = {\bf k}^{2} /(2m) - \mu$ and 
$q=({\bf q},\Omega_{\nu})$ 
is a four-momentum, with wave vector 
${\bf q}$ and Matsubara frequency $\Omega_{\nu}$ ($\nu$ integer).
In an analogous way, one can show that in the particle-particle channel the 
contributions of the vertex corrections 
and of the two-particle effective interactions other than the rung 
vanish for our choice of the potential.

It is thus evident from these considerations that, with our choice of the 
fermionic interaction, in the strong-coupling limit the 
\emph{skeleton structure\/} of the diagrammatic 
theory can 
be constructed only with the particle-particle ladder (\ref{p-p ladder}) plus 
an infinite number of interaction vertices.
A careful diagrammatic analysis considered in detail in Ref.~\onlinecite{PS00}
then shows that:
(a) Bare composite bosons correspond to fermionic particle-particle ladders;
(b) The interaction among bare composite bosons is described by 4-point, 
6-point vertices, and so on, which correspond to the product of $4,6,\ldots$, 
fermionic bare Green's functions (with one internal four-momenta integration).
The correspondence rules for the bosonic Green's function and the 4-point 
vertex are shown in Fig. 2.

\begin{figure}
\narrowtext
\epsfxsize=3.3in 
\epsfbox{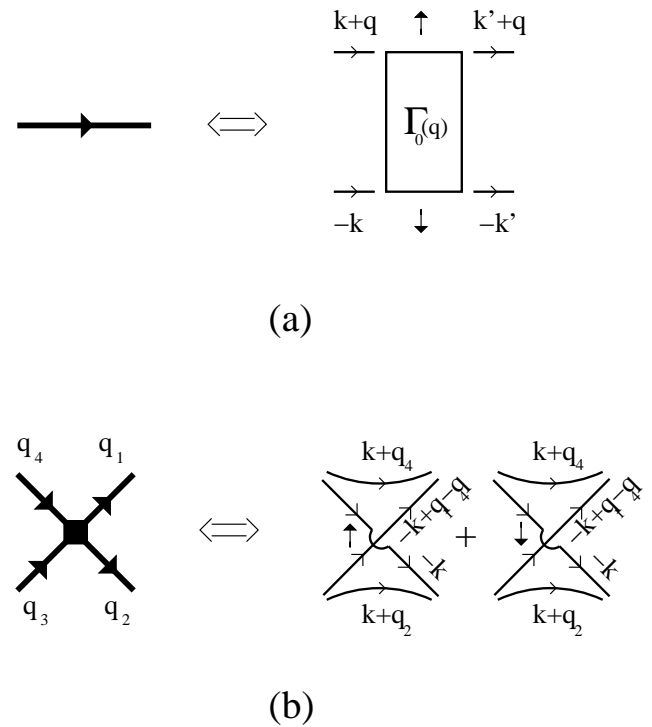}
\vspace{.2truecm}
\caption{ Graphical correspondence (a) between the 
bare propagator for composite bosons (represented by a thick line) and the
fermionic particle-particle ladder, and (b)
between the effective two-boson interaction and the four-point vertex.}        \label{fig:cobos}
\end{figure}

In particular, in the strong-coupling limit (whereby $\beta |\mu| \gg 1$),\cite{footnote-mu} the particle-particle 
ladder $\Gamma_0(q)$ has the following 
\emph{polar structure\/}:\cite{Haussmann}
\begin{equation}
\Gamma_0(q)\approx-\frac{4 \pi}{m^{2} a_{F}}\frac{ 1 + 
\sqrt{1 + \left(-i\Omega_{\nu} +
\frac{{\bf q}^{2}}{4 m} - \mu_{B}\right)\epsilon_{0}^{-1}}} 
{i\Omega_{\nu} - \left(\frac{{\bf q}^{2}}{4m} - \mu_{B}\right)}
\label{pp-sc}
\end{equation}
where we have used the definition $\mu_{B} = 2\mu + \epsilon_{0}$ for the 
bosonic chemical potential ($\epsilon_0=1/(m a_F^2)$ being the bound-state 
energy of the fermionic two-body problem). Note that
(apart from the residue being different from unity) the expression 
(\ref{pp-sc}) resembles a ``free'' boson propagator 
with mass $2 m$.
 The (four-point) \emph{effective two-boson interaction\/} reads instead
\begin{eqnarray}   
& &\tilde{u}_{2}(q_{1} \dots q_{4}) \, = \, \delta_{q_1+q_2,q_3+q_4} 
\left(\frac{{\cal V}}{\beta}\right)^{2}  \frac{2}{\beta {\cal V}}
\label{two-body-potential}\\
& &\times \sum_k 
{\cal G}^0(-k){\cal G}^0(k+q_2){\cal G}^0(-k+q_1-q_4){\cal G}^0(k+q_4)
\nonumber
\;\; .
\end{eqnarray}
Here ${\cal G}^{0}(k) = [i\omega_n -{\bf k}^2/(2m)-\mu]^{-1}$ is a bare 
fermionic Green's function ($\omega_n$ being a fermionic Matsubara frequency).
Note that the interaction (\ref{two-body-potential}) depends on wave vectors 
as well as on Matsubara frequencies, revealing in this way the composite 
nature of the bosons. 
The factor of $2$ in the definition (\ref{two-body-potential}) 
corresponds
to the two different sequences of spin labels that can be attached to the four
fermionic Green's functions, as shown graphically in Fig.~\ref{fig:cobos}(b). 

In the strong-coupling limit $\beta |\mu| \gg 1$, a typical 
value of the two-boson effective interaction is obtained by 
setting $q_1 = \cdots = q_4 = 0$ in Eq.~(\ref{two-body-potential}). One gets:
\begin{equation}
\tilde{u}_{2}(0) \, = \, 2 \, \left( \frac{{\cal V}}{\beta} \right)^{2}
    \left( \frac{m^{2} a_{F}}{8 \pi} \right)^{2} \, u_{2}(0)  \label{u2-tilde-limit}
\end{equation}
where \cite{PS-96}
\begin{equation}  
u_{2}(0) \, = \, \frac{4 \pi (2a_{F})}{2m} \, .                \label{u2-limit}
\end{equation}
The factor $m^{2} a_{F}/(8 \pi)$ in Eq.~(\ref{u2-tilde-limit}) reflects the difference 
between the true bosonic propagator and the particle-particle ladder in the 
strong-coupling limit [cf. also Eq.~(\ref{pp-sc})]. 
Owing to this difference, $u_{2}(0)$ given by Eq.~(\ref{u2-limit}) (and not
$\tilde{u}_{2}(0)$ given by Eq.~(\ref{u2-tilde-limit})) has to be 
identified with the boson-boson interaction at zero 
four-momenta.
Note that $u_{2}(0)$ is \emph{positive\/} in the strong-coupling limit, thus 
ensuring the \emph{stability\/} of the bosonic system.

Recall further that the scattering length $a^{\mathrm{Born}}_{B}$ within 
the Born 
approximation, obtained for a pair of true bosons (each of mass $2m$) mutually 
interacting via a two-body potential with Fourier transform $u_{2}(0)$ at 
zero wave 
vector, is given by $a^{\mathrm{Born}}_{B} = 2m u_{2}(0)/(4 \pi)$. 
Equation~(\ref{u2-limit}) then
yields the following relation between the bosonic and fermionic scattering 
lengths within the Born approximation:
\begin{equation}
a^{\mathrm{Born}}_{B} \, = \, 2 \, a_{F} \, .                             
\label{a-Born}
\end{equation}
The result (\ref{a-Born}) was also obtained in Ref.~\onlinecite{Haussmann}
within the fermionic self-consistent T-matrix approximation (which corresponds to the bosonic 
Hartree-Fock approximation in the strong-coupling
limit). In that reference, the result (\ref{a-Born}) was erroneously regarded 
to be the full value of the 
scattering length $a_{B}$ for a ``dilute'' system of composite bosons. 
We will, in fact,  show in the next section that the result (\ref{a-Born}) 
differs from 
$a_{B}$ when \emph{all\/} bosonic T-matrix diagrams 
are properly taken into account.

Besides the four-point vertex (\ref{two-body-potential}), the composite nature 
of 
the bosons produces (an infinite set of) additional vertices.
One can show that 
all interaction vertices can be neglected in comparison with the 
four-point vertex in the strong-coupling limit.\cite{PS00} 
In this limit, one can thus construct all diagrams representing the 
two-particle Green's
function in the particle-particle channel in terms only of the bare 
ladder 
and of the four-point interaction vertex.
This is precisely what one would expect on physical grounds, since the 
interactions 
involving more than two bodies become progressively less effective as the 
composite 
bosons overlap less when approaching the strong-coupling limit.

\section{Numerical results for the scattering length of composite bosons}

In three dimensions the \emph{scattering length\/} $a$ characterizes the 
low-energy collisions for the scattering from an ordinary potential.
For the mutual scattering of two particles (each of mass $M$), $a$ can be 
expressed 
by the relation $t(0) = 4\pi a/M$ in terms of the ordinary T-matrix $t(0)$ in 
the limit of vanishing wave vector.
In particular, within the Born approximation $t(0)$ is replaced by the
Fourier trasform $u(0)$ of the interparticle potential for vanishing wave 
vector.

\begin{figure}
\narrowtext
\epsfxsize=3.3in 
\epsfbox{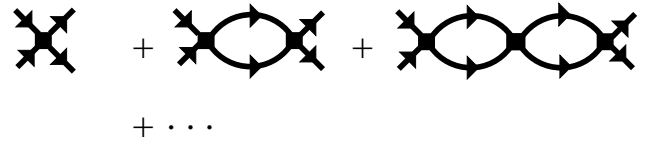}
\vspace{.2truecm}
\caption{T-matrix diagrams for the composite bosons. Heavy lines and squares
are defined as in Fig.~2.}         
\label{fig:comboslo}
\end{figure}

For composite bosons, the T-matrix 
$\bar{t}_{B}(q_{1},q_{2},q_{3},q_{4})$ is defined by the following integral 
equation (cf.~Fig.~\ref{fig:comboslo}):
\begin{eqnarray}
& &\bar{t}_{B}(q_{1},q_{2},q_{3},q_{4}) =\bar{u}_{2}(q_{1},q_{2},q_{3},q_{4}) 
\nonumber\\ 
& &-\frac{1}{\beta {\cal V}} \, \sum_{q_{5}} \,
 \bar{u}_{2}(q_{1},q_{2},q_{5},q_{1}+q_{2}-q_{5}) 
\Gamma_{0}(q_{5}) \, \Gamma_{0}(q_{1}+q_{2}-q_{5})\nonumber\\
& &\times \bar{t}_{B}(q_{1}+q_{2}-q_{5},q_{5},q_{3},q_{4}) 
\label{bosonic-t-matrix}
\end{eqnarray}
where $\bar{u}_{2}$ is proportional to the effective two-boson
interaction of Eq.~(\ref{two-body-potential}):
\begin{eqnarray}
& &\bar{u}_{2}(q_{1} \dots q_{4}) \, = \,  \frac{1}{\beta {\cal V}} \, 
\sum_{k} \nonumber\\
& &\times
{\cal G}^0(-k){\cal G}^0(k+q_2){\cal G}^0(-k+q_1-q_4){\cal G}^0(k+q_4) \; .
\label{two-body-potential-bar}     
\end{eqnarray}

Similarly to the problem of point-like bosons, we \emph{define\/} the 
scattering length $a_{B}$ for the composite 
bosons (each of mass $2m$) in the strong-coupling limit and for vanishing 
density, by setting $t_{B}(0) = 4 \pi  a_{B}/(2m)$.  
Here $t_{B}(0) = (8 \pi/(m^{2} a_{F}))^{2} \, \bar{t}_{B}(0)$ and
$\bar{t}_{B}(0) \equiv \bar{t}_{B}(0,0,0,0)$. The rescaling between 
$t_{B}(0)$ and $\bar{t}_{B}(0)$  is due to the difference between the 
true bosonic propagator and the particle-particle ladder in the 
strong-coupling limit. The same rescaling is consistently used when defining 
the boson-boson interaction [cf. Eqs.~(\ref{u2-tilde-limit}) and
(\ref{two-body-potential})].

To the lowest order in the effective interaction between the composite bosons,
we can replace $t_{B}(0)$ by $u_{2}(0)$ and write $u_{2}(0)= 4 
\pi a^{\mathrm{Born}}_{B}/(2m)$, within the Born approximation.
Comparison with Eq.~(\ref{u2-limit}) yields then the value 
$a^{\mathrm{Born}}_{B} = 2 \, a_{F}$, as anticipated by Eq.~(\ref{a-Born}).
 
In order to obtain the exact value of $\bar{t}_{B}(0)$ (and hence of the 
scattering length $a_B$), it is 
convenient to determine first the auxiliary
quantity $\bar{t}_{B}(q,-q,0,0)$ by solving the following \emph{closed-form\/} equation 
\begin{eqnarray}
& &\bar{t}_{B}(q,-q,0,0) = \bar{u}_{2}(q,-q,0,0) \label{bosonic-t-matrix-0}\\ 
& & - \frac{1}{\beta {\cal V}} \, \sum_{q'} \,
     \bar{u}_{2}(q,-q,q',-q')\Gamma_{0}(q') \, \Gamma_{0}(-q') \,
     \bar{t}_{B}(q',-q',0,0)   \,\, , \nonumber 
\end{eqnarray}
obtained from Eq.~(\ref{bosonic-t-matrix}) by setting $q_{1}=-q_{2}=q$ and
$q_{3}=q_{4}=0$.
This integral equation can be solved by standard numerical methods, e.g., by
reverting it to the solution of a system of coupled linear equations.
Note that, since we are here interested in the calculation of a two-boson 
quantity, the zero-density limit has to be taken in 
Eq.~(\ref{bosonic-t-matrix-0}).
This implies that $\mu_B=0$ (or equivalently $2\mu=\epsilon_0$) and 
$T=0$.\cite{footnote-mu} The discrete bosonic frequency $\Omega_{\nu}$ 
becomes a continuous variable accordingly.

Numerical calculation of Eq.~(\ref{bosonic-t-matrix-0}) requires 
us to introduce a finite-size mesh for the variables ($|{\bf q}|$, $\Omega$) 
and ($|{\bf q}'|$, $\Omega'$), with the angular integral over $\hat{q'}$
affecting only the function $\bar{u}_{2}(q,-q,q',-q')$.
Equation (\ref{bosonic-t-matrix-0}) is thus reduced to a set of coupled 
equations
for the unknowns $\bar{t}_{B}(|{\bf q}|,\Omega;|{\bf q}|,-\Omega;0;0)$,
which were solved by the Newton-Ralphson algorithm with a linear
interpolation for the integrals over $|{\bf q}'|$ and $\Omega'$.
In this way, we are led to the result
\begin{equation}
a_{B} = 0.75 \, a_F
\end{equation}
within an estimated $5\%$ numerical accuracy.

To summarize, we have shown that, in the strong-coupling limit of the 
fermionic attraction, the value
$a_{B} = 0.75 \, a_{F}$ is obtained by a correct summation 
of the bosonic T-matrix diagrams for the composite bosons which form as
bound-fermion pairs. 
This theoretical
prediction could be tested against the experimental data with ultracold Fermi
atoms in a trap, when bosonic molecules are obtained by a Feshbach
resonance.



\begin{references}
\bibitem{footnote-mu} A precise definition of what is meant by  
strong coupling is as follows.
In this limit, the binding energy 
$\epsilon_{0}$ is much larger than the density  
$\rho$ and temperature $T$. Under this condition, the fermionic chemical 
potential approaches the value $-\epsilon_{0}/2$ and 
$\beta |\mu| \gg 1$.
\end{references}
\end{document}